\documentclass{article}

\usepackage{arxiv}

\usepackage[utf8]{inputenc} 
\usepackage[T1]{fontenc}    
\usepackage[colorlinks]{hyperref}       
\usepackage{url}            
\usepackage{booktabs}       
\usepackage{amsfonts}       
\usepackage{nicefrac}       
\usepackage{microtype}      
\usepackage{dirtytalk}
\usepackage{xcolor}
\usepackage{soul}
\usepackage{array}
\usepackage{stmaryrd}
\newcolumntype{P}[1]{>{\centering\arraybackslash}p{#1}}
\include{javalisting}
\usepackage{graphicx}

\title{COVID-19 Lung Lesion Segmentation Using a Sparsely\\Supervised Mask R-CNN on Chest X-rays Automatically Computed from Volumetric CTs}

\author{
  Vignav Ramesh$^{1, 2}$\thanks{Corresponding author: \href{mailto:rvignav@gmail.com}{\texttt{rvignav@gmail.com}}}\hspace{0.4em}, Blaine Rister$^{3}$, Daniel L. Rubin$^{4}$ \\\\
   	$^{1}$ Saratoga High School, 20300 Herriman Ave, Saratoga, CA 95070, USA\\
	$^{2}$ Department of Biomedical Data Science, Stanford University, 1265 Welch\\Road,
Stanford, CA 94305, USA \\
$^{3}$ Department of Electrical Engineering, Stanford University, 350 Jane\\Stanford Way, Stanford, CA 94305, USA\\
   $^{4}$ Professor of Biomedical Data Science, Radiology, and Medicine (Biomedical Informatics),\\Stanford University, Room X-335, 1265 Welch Road,
Stanford, CA 94305, USA

}

\begin{document}
\maketitle

\begin{abstract}
Chest X-rays of coronavirus disease 2019 (COVID-19) patients are frequently obtained to determine the extent of lung disease and are a valuable source of data for creating artificial intelligence models. Most work to date assessing disease severity on chest imaging has focused on segmenting computed tomography (CT) images; however, given that CTs are performed much less frequently than chest X-rays for COVID-19 patients, automated lung lesion segmentation on chest X-rays could be clinically valuable. There currently exists a universal shortage of chest X-rays with ground truth COVID-19 lung lesion annotations, and manually contouring lung opacities is a tedious, labor-intensive task. To accelerate severity detection and augment the amount of publicly available chest X-ray training data for supervised deep learning (DL) models, we leverage existing annotated CT images to generate frontal projection \say{chest X-ray} images for training COVID-19 chest X-ray models. In this paper, we propose an automated pipeline for segmentation of COVID-19 lung lesions on chest X-rays comprised of a Mask R-CNN trained on a mixed dataset of open-source chest X-rays and coronal X-ray projections computed from annotated volumetric CTs. On a test set containing 40 chest X-rays of COVID-19 positive patients, our model achieved \textsc{IoU} scores of $0.81 \pm  0.03$ and $0.79 \pm  0.03$ when trained on a dataset of 60 chest X-rays and on a mixed dataset of 10 chest X-rays and 50 projections from CTs, respectively. Our model far outperforms current baselines with limited supervised training and may assist in automated COVID-19 severity quantification on chest X-rays.
\end{abstract}


\section{Introduction}
Coronavirus disease 2019 (COVID-19), a febrile respiratory illness caused by severe acute respiratory syndrome coronavirus 2 (SARS-CoV-2), was initially reported to the World Health Organization (WHO) in December 2019.$^1$ As of February 14, 2021, the WHO reported 108,153,741 worldwide cases and 2,381,295 confirmed deaths.$^{2,3}$ Chest X-rays are frequently obtained to determine the extent of lung disease in potential COVID-19 patients and are a valuable source of data for creating artificial intelligence models for both COVID-19 prediction and severity quantification.

Findings from the COVID-19 Immune Response Study suggest that early intervention is critical for COVID-19 treatment; patients with moderate illness have not yet developed end-organ damage, meaning that treatment is more effective.$^{4}$ Thus, clinicians currently utilize a variety of methods for early COVID-19 detection. Reverse transcription polymerase chain reaction (RT-PCR) is a standard diagnostic method that involves the extraction of nucleic acid from samples obtained by oropharyngeal swab, nasopharyngeal swab, bronchoalveolar lavage, or tracheal aspirate.$^{5}$ However, RT-PCR does not provide insight into the severity of COVID-19 lung infection; severity quantification is a task that image analysis is best suited to perform.{$^{6,7}$} Computed tomography (CT) imaging has been used to identify areas of consolidation or ground glass opacities in the lungs$^{8}$ and thus has value in visualizing the extent of COVID-19 infection (\href{https://arxiv.org/abs/2004.10987}{Yan et al., 2020}; \href{https://arxiv.org/abs/2004.14133}{Fan et al., 2020}; \href{https://www.ncbi.nlm.nih.gov/pmc/articles/PMC7605758/}{Oulefki et al., 2020}; \href{https://www.researchsquare.com/article/rs-40406/v1.pdf}{Akbari et al., 2020}). Most work to date assessing disease severity on chest imaging has focused on segmenting CT images; however, given that CT scans are performed much less frequently than chest X-rays for COVID-19 patients, automated lung lesion segmentation and severity quantification on chest X-rays could be clinically valuable. There currently exists a universal shortage of chest X-rays with ground truth COVID-19 lung lesion annotations, and manually contouring lung opacities on chest X-rays is a tedious and labor-intensive task. However, CT imaging is a modality that has been commonly obtained in countries other than the United States (namely China, Russia, and India), and thus, CTs with ground truth lung lesion annotations are publicly available.$^{9}$ To accelerate diagnosis and severity detection, increase access to treatment for a wider demographic of COVID-19 patients, and augment the amount of open-source chest X-ray training data for supervised deep learning (DL) models, an automated method of segmenting lesions on chest X-rays of COVID-19 patients that utilizes publicly available CT data is critically needed.

\subsection{Prior Work}

Few other published methods of segmenting COVID-19 lung lesions on chest X-rays exist. Tang, Sun, and Li proposed a U-Net with a ResNet-18 backbone for segmentation of opacity regions on chest X-rays.$^{11}$ While the model detected lung lesions fairly accurately, high-density anatomical structures such as bronchial trees were often perceived as additional opacity regions and thereby compromised model predictions, indicating that the network leaves room for improvement. Oh, Park, and Ye proposed a patch-based convolutional neural network approach with a relatively small number of trainable parameters for COVID-19 diagnosis inspired by statistical analysis of potential imaging biomarkers of chest X-ray radiographs.$^{12}$ They adopted an extended fully convolutional DenseNet103 with a ResNet-18 backbone for the segmentation task; however, this model only segments the lungs and heart on chest X-rays and does not provide any information on opacity regions beside saliency maps, which do not contain accurate lesion outlines and thereby are often difficult to interpret for clinical purposes.

\subsection{Contributions}

We develop an automated pipeline for COVID-19 lung lesion segmentation on chest X-rays. Due to the lack of publicly available annotated chest X-ray data, we implement a pixel-based algorithm (a method operating at the pixel level) that generates coronal X-ray projections from annotated volumetric CTs to augment the training dataset. A Mask R-CNN framework is then trained on this mixed dataset. Our model achieves superior accuracy with only limited supervised training.

The main contributions of the paper can be summarized as follows:
\begin{enumerate}
    \item We present the first publicly available, open-source chest X-ray dataset containing over 100 images, assembled from various public sources (see Section 2.1), with COVID-19 lung lesion annotations produced by our Mask R-CNN model.
    \item We implement a pixel-based algorithm to compute a coronal X-ray projection (with overlaid segmentations) from an annotated CT volume.
    \item We implement a Mask R-CNN$^{13}$ architecture for automated segmentation of COVID-19 lung lesions on chest X-rays.
    \item We demonstrate the potential for mixed datasets (chest X-rays as well as projections from CTs) to improve the performance of deep learning models that operate on chest X-rays. Moreover, even with a small training dataset, the proposed model achieves superior accuracy without overfitting, indicating that our approach requires only sparse supervision (in the context of this paper, sparse supervision refers to extremely limited supervised training).
\end{enumerate}

\section{Materials and Methods}
\subsection{Data}
We obtained a total of 241 annotated CT volumes from the following sources:
\begin{itemize}
    \item Kaggle$^{14-16}$ (20 cases): Compilation of 10 Radiopedia volumes and 10 Coronacases volumes
    \item MosMedData$^{17}$ (22 cases): Obtained between March 1, 2020 and April 25, 2020 and provided by medical hospitals in Moscow, Russia
    \item COVID-19-20 Lung CT Lesion Segmentation Grand Challenge$^{18-22}$ (199 cases): Unenhanced chest CTs from 199 patients with positive RT-PCR for SARS-CoV-2
\end{itemize}

 We also obtained 100 chest X-rays with ground truth lung lesion annotations from the following source:
\begin{itemize}
    \item General Blockchain Inc.$^{23-25}$: Compiled from a variety of public sources as well as through indirect collection from hospitals and physicians
\end{itemize}

\subsection{CT to X-ray Conversion}

\includegraphics[width=\textwidth]{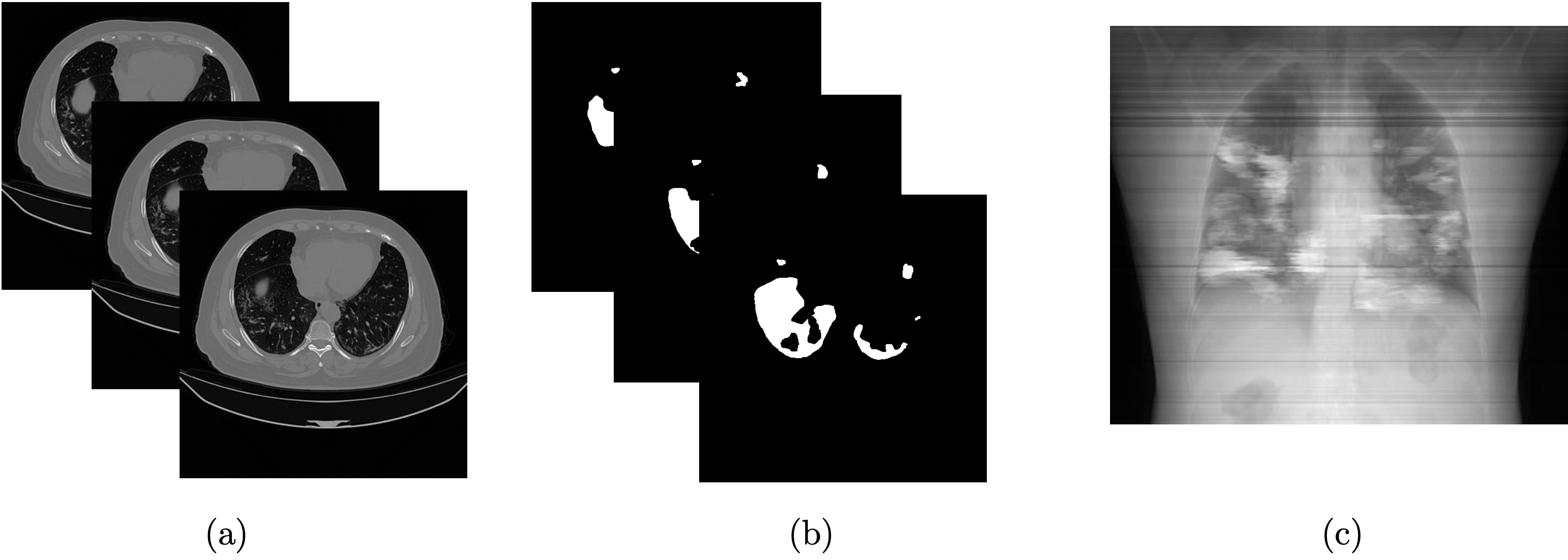}
\textbf{Figure 1: CT to X-ray conversion.} The proposed algorithm uses (a) a volumetric CT and (b) a volume of mask slices to compute (c) an annotated coronal X-ray.

Generating projections from volume data is a well-known task (\href{https://pubs.rsna.org/doi/full/10.1148/rg.255055044}{Dalrymple et al., 2005}; \href{https://ucdavis.pure.elsevier.com/en/publications/value-of-axial-and-coronal-maximum-intensity-projection-mip-image}{Valencia et al., 2006}; \href{https://www.sciencedirect.com/science/article/pii/S0378603X16301759}{Özkan et al., 2016}). We implement a pixel-based re-projection method, modeled as a sub-problem of ray tracing as outlined in [26], to compute chest X-rays as coronal projections of volumes of axial CT slices. Equation (1) summarizes the proposed method:
\begin{equation}
    \Theta(x,z) = \sum_{i=1}^{Y} \Phi(x,i,z) \quad \quad \forall \: (x,z) \in \llbracket1,X\rrbracket \times \llbracket1,Z\rrbracket,
\end{equation}
where $\Phi$ is an $X \times Y \times Z$ 3D array denoting the volumetric CT and $\Theta$ is an $X \times Z$ matrix denoting the computed X-ray.

Besides being used to compute coronal X-ray projections, the CT to X-ray conversion algorithm is also used to generate the labels for each sample in the training dataset. Given a volume of ground truth mask slices in grayscale format, where a nonzero pixel value is considered part of an opacity region and vice versa, the CT to X-ray conversion algorithm is used to generate a coronal mask projection from the axial mask volume. A recursive floodfill$^{27}$ is then performed on the mask projection to generate sets of pixel values, where each set stores the coordinates of the pixel values within a single disjoint opacity region on the projection. Each set is then used to generate a concave hull$^{28,29}$, a data structure that computes a boundary polygon from a list of points by composing the edges of a series of triangles constructed from randomly chosen trios of points. After the coronal mask projection is constructed, each of its opacity regions has a corresponding concave hull that stores the boundary points of its representative polygon. These concave hulls are passed to the Mask R-CNN as labels for the X-ray projections in the training dataset.

\subsection{Lung Lesion Segmentation}

We employ a naive implementation of the Mask R-CNN framework for the task of instance segmentation.$^{30}$ In a Mask R-CNN architecture, training samples are fed into a ResNet-101 backbone network (see Figure 2), convolved, and passed to the Region Proposal Network (RPN) to generate a set of proposed regions possibly containing lung lesions. Anchors corresponding with each region of interest are then passed through a series of feature maps to generate masks outlining COVID-19 lung lesions on the input chest X-ray. Object classes and bounding boxes are computed via a series of fully connected layers. The task of COVID-19 lung lesion segmentation is posed as a problem of binary classification between the image background and lung lesions. The final output is a predicted mask corresponding with the input chest X-ray, which can then be overlaid on the input image for clinical use. Figure 3 provides a diagram of this network architecture.

{\begin{center} \includegraphics[width=0.8\textwidth]{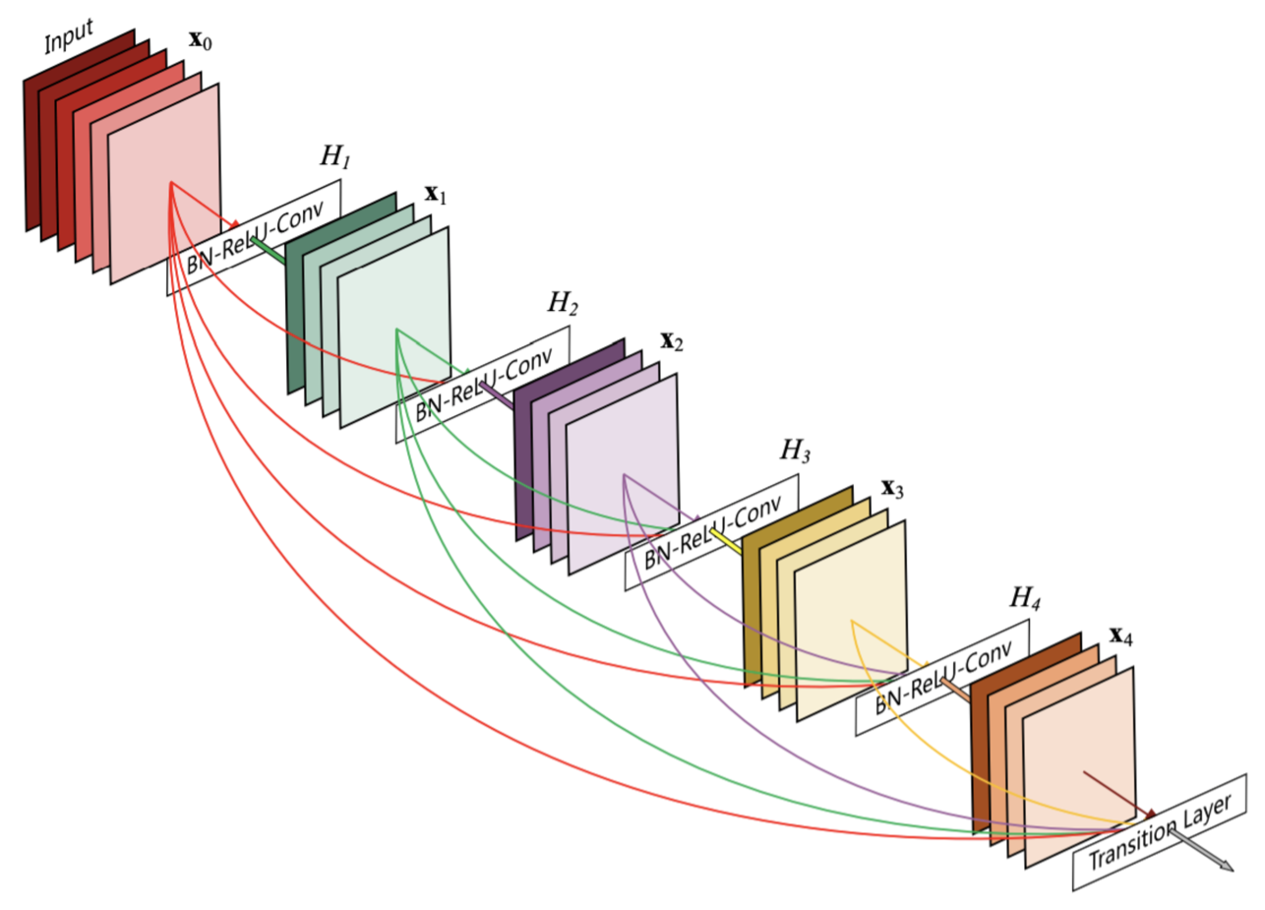} \end{center}}
\textbf{Figure 2: ResNet-101 backbone network architecture.$^{31}$} The backbone network is structured as a Feature Pyramid Network (FPN)$^{32}$-style deep neural network consisting of a bottom-up pathway, top-down pathway, and lateral connections (convolutional operations between corresponding levels of the two pathways).$^{33}$

\includegraphics[width=\textwidth]{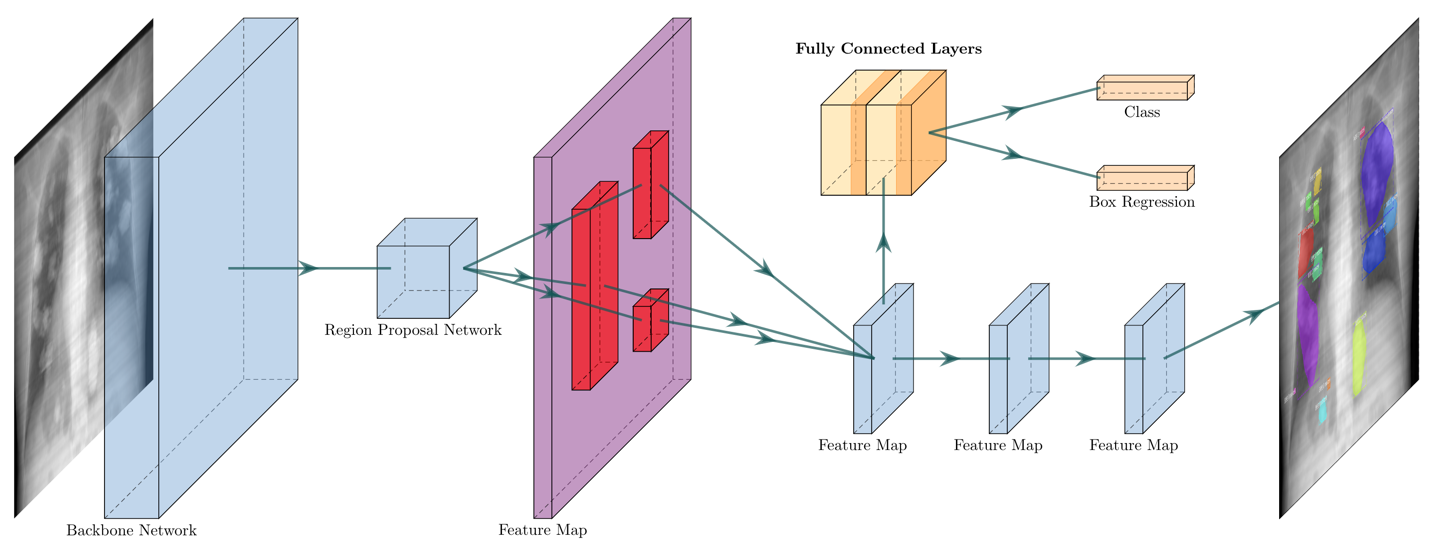}
{\textbf{Figure 3: Mask R-CNN network architecture.} Training images are fed into a backbone network (see Figure 2), which then passes a network representation of the training samples to the Region Proposal Network (RPN). The RPN scans the top-bottom pathway of the backbone network and proposes regions which may contain objects of interest on the feature map. RPN anchors are then fed into a series of feature maps, allowing for the parallel execution of two operations: 1) the creation of masks; and 2) the generation of object classes and bounding boxes using a series of fully connected layers. }

\subsubsection{Default Setting}

We implemented the Mask R-CNN architecture using TensorFlow and Keras. We trained for 30 epochs, each with 200 training steps and 50 validation steps, at batch size 2. We used ResNet-101 as the base encoder network with backbone strides of 4, 8, 16, 32, and 64 and a top-down pyramid size of 256. For the losses, we used RPN class loss, RPN bounding box loss, Mask R-CNN class loss, Mask R-CNN bounding box loss, and Mask R-CNN mask loss. We used a gradient clip norm of 5.0, an image shape of 1024 × 1024 × 3, a learning momentum of 0.9, a learning rate of 0.001, a weight decay of 0.0001, and a mask shape of 28 × 28. For the RPN specifications, we used anchor ratios of 0.5, 1, and 2; anchor scales of 32, 64, 128, 256, and 512; an anchor stride of 1; a ROI positive ratio of 0.33; bounding box standard deviations of 0.1 and 0.2; and 200 training ROIs per image. The maximum number of ground truth instances is 15.

Due to the paucity of publicly available chest X-ray data with COVID-19 lung lesion annotations, we also used data augmentation to improve model accuracy and reduce potential overfitting. The following augmentation techniques were applied to the training samples with the specified probabilities:
\begin{itemize}
    \item Horizontal flip: 0.5
    \item 0-10\% Random crop: 1.0
    \item Small Gaussian blur with randomly chosen $\sigma \in [0,0.5]$: 0.5
    \item Contrast normalization
    \item Per-channel pixel multiplication (multiplication ratios sampled from the interval $[0.8,1.2]$): 0.2
    \item Affine transformation: 1.0
    \begin{itemize}
        \item Scale (ratio sampled from $[0.8,1.2]$ for each axis)
        \item Rotate (degrees of rotation randomly chosen from $[-10,10]$, where negative indicates counterclockwise and positive indicates clockwise)
        \item Shear (degrees of shearing randomly chosen from $[-2,2]$ and applied to all image axes)
    \end{itemize}
\end{itemize}

\subsection{Training and Testing Protocol}

We trained identical Mask R-CNN architectures (see Section 2.3) on two different training datasets of the same size for exactly 30 epochs with the same test dataset, allowing us to compare results across training datasets.

\textbf{Dataset 1 (X-rays Only):} The first training dataset consisted of 60 chest X-rays with ground truth annotations obtained from General Blockchain Inc.’s public dataset via random selection.

\textbf{Dataset 2 (Mixed):} The second training dataset consisted of 10 chest X-rays randomly selected from the 60 chest X-rays in Dataset 1 as well as 50 randomly selected X-ray projections from CT volumes obtained from Kaggle, MosMedData, and the COVID-19-20 Lung CT Lesion Segmentation Grand Challenge (see Section 2.1).

\textbf{Test Dataset:} Our model was evaluated on the remaining 40 chest X-rays of COVID-19 positive patients from General Blockchain Inc’s dataset.

\section{Results}

To evaluate our model’s performance on the test set (see Section 2.4), we use the Intersection over Union (\textsc{IoU}) metric. \textsc{IoU} is a similarity metric between the ground truth and the prediction segmentations.$^{34}$

\begin{tabular}{ |P{2.8cm}|P{3.1cm}|P{2.7cm}|P{3.1cm}|P{2.7cm}|  }
  \cline{2-5}
    \multicolumn{1}{c|}{} &  \multicolumn{2}{c|}{\textbf{Training Dataset}} &  \multicolumn{2}{c|}{\textbf{Baseline$^{**}$}}\\
\hline
\textbf{Metric} & Dataset 1 (X-rays Only)$^*$ & Dataset 2 (Mixed)$^*$ & Dataset 1 (X-rays Only)$^*$ & Dataset 2 (Mixed)$^*$\\
\hline

Intersection over Union (\textsc{IoU}) & $0.8056 \pm  0.0266$ & $0.7937 \pm 0.0291 $ & $0.3824 \pm 0.0349$ & $0.4870 \pm 0.0322$  \\
\hline
\end{tabular}\\
\indent $^*$Margins of error obtained via a 1-sample $t$-test for population mean ($\mu$) with 95\% confidence 

\indent $^{**}$Tang, Sun, and Li’s U-Net segmentation model

  When trained on Dataset 1 (X-rays only), the proposed model achieved an \textsc{IoU} score of $0.81 \pm  0.03$. When trained on Dataset 2 (mixed), it achieved an \textsc{IoU} score of $0.79 \pm  0.03$. The similarity between these results indicates that we can replace more than 83\% of chest X-ray training images with X-ray projections generated from CTs while maintaining model accuracy. The following figures contain representative results (see Figure 4 for ground truth and predicted masks; see Figure 5 for predicted segmentations overlaid on chest X-rays from the test set).
 
\includegraphics[width=\textwidth]{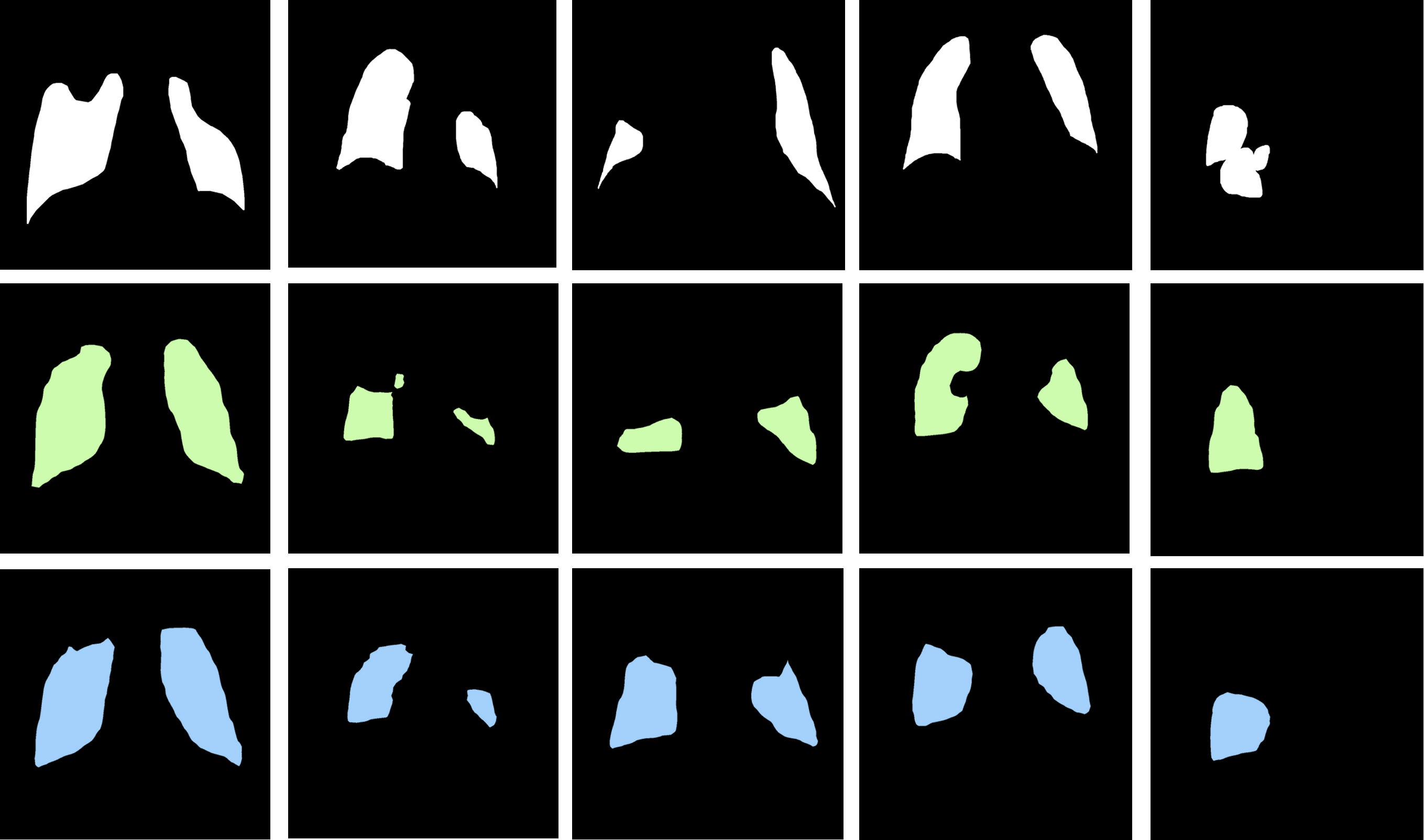}
{\textbf{Figure 4: Ground truth and predicted masks.} Sample of five chest X-rays from the test dataset. The top row (white) displays the ground truth masks, the middle row (green) contains the masks predicted by the model after training on Dataset 1, and the bottom row (blue) contains the masks predicted by the model after training on Dataset 2. Chest X-rays are the same within individual columns.
}

\includegraphics[width=\textwidth]{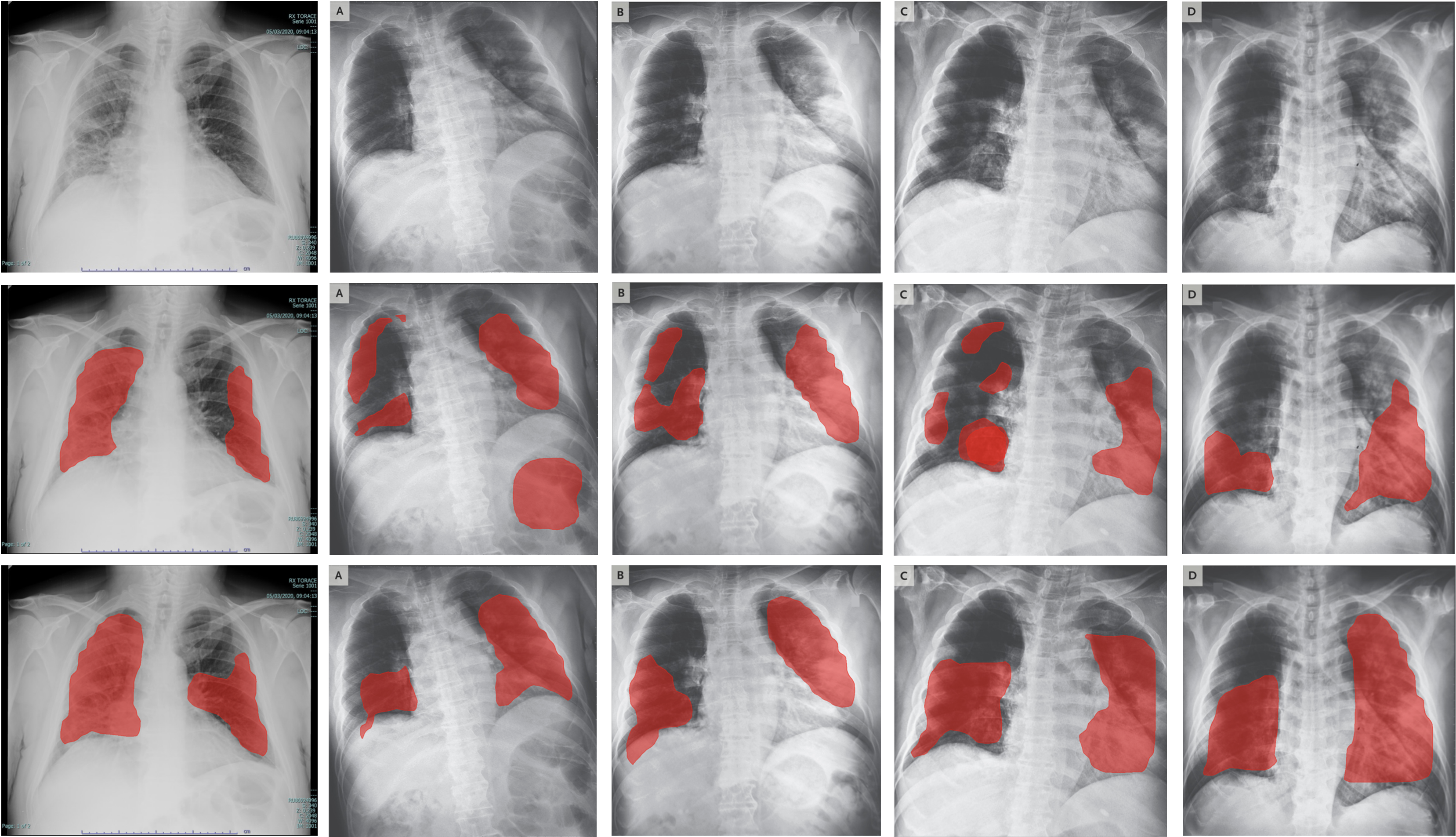}
{ \textbf{Figure 5: Overlaid segmentations.} Sample of five chest X-rays from the test dataset. The top row displays the original X-rays, the middle row contains the predicted segmentations overlaid on the X-rays after training on Dataset 1, and the bottom row contains the predicted segmentations overlaid on the X-rays after training on Dataset 2. Chest X-rays are the same within individual columns.}

\section{Discussion}
  The above results far exceed the few existing published baselines. For instance, Tang, Sun, and Li’s U-Net segmentation model described in Section 1 (the only published COVID-19 lung lesion segmentation framework with publicly available model schematics), achieved \textsc{IoU} scores of $0.38 \pm 0.03$ and $0.49 \pm 0.03$, respectively, both of which are significantly lower than our model's corresponding \textsc{IoU} scores of $0.81 \pm  0.03$ and $0.79 \pm 0.03$. Since we trained and tested our model and the baseline model on the same datasets, our Mask R-CNN likely outperformed Tang, Sun, and Li's U-Net segmentation architecture due to its structure as a series of recurring feature maps rather than contracting and expansive paths, the presence of the RPN, and its greater complexity in the form of a ResNet-101 backbone rather than a ResNet-18 backbone. (Note that in the domain of biomedical image segmentation, it is widely accepted that Mask R-CNN models are more robust than U-Net models.$^{35,36}$) Furthermore, when the predicted masks are on the input images, our results can be utilized in clinical contexts to quantify the amount of lung disease on chest X-rays of COVID-19 patients and can thus assist in prognosis detection and treatment determination.

\section{Conclusions}
We propose a fully automated pipeline to segment COVID-19 lung opacities on chest X-rays. The majority of current work regarding COVID-19 opacity segmentation has been in CT imaging; however, since chest X-rays are much more commonly obtained than CTs, COVID-19 lung disease segmentation on chest X-ray images could enable the development of artificial intelligence applications that use these images to assess disease severity and evaluate or predict progression. Our model first utilizes a pixel-based summation algorithm to compute a coronal X-ray projection from axial CT and mask volumes. A floodfill- and hull-based approach is then employed to generate labels for each chest X-ray projection, which are used to train a Mask R-CNN for the task of instance segmentation. Our model achieved an \textsc{IoU} score of $0.81 \pm  0.03$ when trained on Dataset 1, and an \textsc{IoU} score of $0.79 \pm  0.03$ when trained on Dataset 2. A small number of chest X-rays compared to the number of frontal projections was chosen for Dataset 2 in order to demonstrate the potential for X-ray projections from CT volumes to replace actual chest X-rays as training samples for deep learning models.


  A limitation of our study is that we used small amounts of publicly available data; however, our results still suggest that improved accuracy can be obtained by augmenting chest X-ray data with large numbers of frontal projections of public CT volumes. Training and testing our model on larger datasets could improve future results.

\section*{References}

\begin{enumerate}
    \item Yan Q, et al.: COVID-19 Chest CT Image Segmentation -- A Deep Convolutional Neural Network Solution. arXiv:2004.10987 [cs.EESS], April 2020
\item WHO Coronavirus Disease (COVID-19) Dashboard. Available at https://covid19.who.int. \\Accessed 14 February 2020. 
\item Coronavirus Disease 2019 (COVID-19) - Symptoms and Causes. Available at \\https://www.mayoclinic.org/diseases-conditions/coronavirus/symptoms-causes/syc-20479963. \\Accessed 25 December 2020. 
\item Can Early Intervention Slow the Progression of COVID-19? Available at \\https://www.providence.org/news/uf/637014517. Accessed 25 December 2020.
\item Bai HX, et al.: Performance of Radiologists in Differentiating COVID-19 from Non-COVID-19 Viral Pneumonia at Chest CT. Radiology 296:E46-E54, 2020
\item Ai T, et al.: Correlation of Chest CT and RT-PCR Testing for Coronavirus Disease 2019 (COVID-19) in China: A Report of 1014 Cases. Radiology 296:E32-E40, 2020
\item Fang Y, Zhang H, Xie J, Lin M, Ying L, Pang P, Ji W: Sensitivity of Chest CT for COVID-19: Comparison to RT-PCR. Radiology 296:E115-E117, 2020
\item Wang D, et al.: Clinical Characteristics of 138 Hospitalized Patients With 2019 Novel Coronavirus-Infected Pneumonia in Wuhan, China. JAMA 323:1061-1069, 2020
\item Ozsahin I, et al.: Review on Diagnosis of COVID-19 from Chest CT Images Using Artificial Intelligence. Comput Math Methods, 2020
\item Romera-Paredes B, Philip HST: Recurrent Instance Segmentation. arXiv:1511.08250 [cs.CV], October 2016
\item Tang H, et al.: Segmentation Model of the Opacity Regions in the Chest X-Rays of the Covid-19 Patients in the US Rural Areas and the Application to the Disease Severity. medRxiv, October 2020
\item Oh Y, et al.: Deep Learning COVID-19 Features on CXR Using Limited Training Data Sets. arXiv:2004.05758 [cs.EESS], May 2020
\item He K, et al.: Mask R-CNN. arXiv:1703.06870 [cs.CV], January 2018
\item Coronacases. Available at https://coronacases.org/. Accessed 25 December 2020.
\item Radiopaedia. Available at https://radiopaedia.org/. Accessed 25 December 2020. 
\item Ma J, Ge C, Wang Y, An X, Gao J, Yu Z, He J.: COVID-19 CT Lung and Infection Segmentation Dataset. Zenodo, April 2020 
\item Morozov SP, Andreychenko AE, Pavlov NA, Vladzymyrskyy AV, Ledikhova NV, Gombolevskiy VA, Blokhin IA, Gelezhe PB, Gonchar AV, Chernina VY: MosMedData: Chest CT Scans With COVID-19 Related Findings Dataset. arXiv:2005.06465 [cs.CY], May 2020
\item Data - Grand Challenge. Available at https://covid-segmentation.grand-challenge.org/Data/. \\Accessed 25 December 2020.
\item An P, et al.: CT Images in COVID-19. TCIA, 2020
\item Clark K, et al.: The Cancer Imaging Archive (TCIA): Maintaining and Operating a Public Information Repository. J Digit Imaging 26:1045-1057, 2013
\item NVIDIA NGC. Available at https://ngc.nvidia.com/catalog/models/nvidia:clara\_train\_covid19\_ct\_\\lesion\_seg. Accessed 25 December 2020.
\item ITK-SNAP Home. Available at http://www.itksnap.org/pmwiki/pmwiki.php. Accessed 25 December 2020.
\item General Blockchain Inc: Covid-19 Chest Xray Segmentations Dataset. Available at https://github.co\\m/GeneralBlockchain/covid-19-chest-xray-segmentations-dataset. Accessed 25 December 2020.
\item Cohen JP, et al.: Predicting COVID-19 Pneumonia Severity on Chest X-Ray with Deep Learning. arXiv:2005.11856 [cs.EESS], June 2020
\item Lin T, et al. Microsoft COCO: Common Objects in Context. arXiv:1405.0312 [cs.CV], February 2015
\item Levoy M: Efficient Ray Tracing of Volume Data. ACM Trans Graph 9:245-261, 1990
\item Burtsev SV, Ye PK: An Efficient Flood-Filling Algorithm. Comput Graph 17:549-561, 1993
\item The Concave Hull. Available at https://towardsdatascience.com/the-concave-hull-c649795c0f0f. Accessed 25 December 2020.
\item Fast Concave Hull Implementation in Python. Available at https://gist.github.com/AndreLester/\\589ea1eddd3a28d00f3d7e47bd9f28fb. Accessed 25 December 2020.
\item Exploring DenseNets and a Comparison with Other Deep Architectures. Available at https://mediu\\m.com/@sraosumanth/exploring-densenets-and-a-comparison-with-other-deep-architectures-85f02\\597400a. Accessed 25 December 2020.
\item Ke L, et al.: Deep Occlusion-Aware Instance Segmentation with Overlapping BiLayers. \\arXiv:2103.12340 [cs.CV], March 2021
\item Lin T, et al.: Feature Pyramid Networks for Object Detection. arXiv:1612.03144 [cs.CV], April 2017
\item Simple Understanding of Mask RCNN. Available at https://alittlepain833.medium.com/simple-understanding-of-Mask R-CNN-134b5b330e95. Accessed 25 December 2020.
\item Generalized Intersection over Union. Available at https://giou.stanford.edu/. Accessed 25 December 2020.
\item Durkee MS, et al.: Comparing Mask R-CNN and U-Net Architectures for Robust Automatic Segmentation of Immune Cells in Immunofluorescence Images of Lupus Nephritis Biopsies. Imaging, Manipulation, and Analysis of Biomolecules, Cells, and Tissues XIX 11647:1-10, 2021
\item Vuola AO, et al.: Mask-RCNN and U-Net Ensembled for Nuclei Segmentation. arXiv:1901.10170 [cs.CV], January 2019
\end{enumerate}
\end{document}